\documentstyle[titlepage,12pt]{article}

\begin{document}
\title{Circular string-instabilities in curved spacetime}
\author{A.L.Larsen\\
Observatoire de Paris, Section de Meudon, Demirm \\
5 Place Jules Janssen, 92195 Meudon Principal Cedex, France\thanks{On
leave of absence from Nordita, Blegdamsvej 17, DK-2100 Copenhagen \O.
\hspace*{7mm}E-mail: allarsen@nbivax.nbi.dk}}
\maketitle
\vskip 12pt
\begin{abstract}
We investigate the connection between curved spacetime and the emergence
of string-instabilities, following the approach developed by Loust\'{o}
and S\'{a}nchez for de Sitter and black hole spacetimes. We analyse
the linearised equations determining
the comoving physical (transverse) perturbations on circular strings
embedded in Schwarzschild, Reissner-Nordstr\"{o}m and de Sitter
backgrounds. In all 3 cases we find that
the "radial" perturbations grow infinitely for $r\rightarrow 0$
(ring-collapse), while the "angular" perturbations are bounded in this
limit. For $r\rightarrow\infty$ we find that the perturbations in both physical
directions (perpendicular to the string world-sheet in 4 dimensions) blow
up in the case of de Sitter space. This confirms results recently
obtained by Loust\'{o} and S\'{a}nchez who considered perturbations around
the string center of mass.
\end{abstract}
\newpage
\section{Introduction}
The classical equations of motion for a string in curved spacetime are
generally non-integrable due to their highly non-linear nature, and even if the
system for some specific background can be shown to be integrable, it may be
a very hard task to actually  write down the general solution in closed form.
In many cases of interest it is on the other hand not so difficult to find
special solutions. The standard way is to make an {\it Ansatz} that somehow
exploits possible symmetries of the background and somehow is also based on
physical insight in the specific case under consideration. If properly
chosen this {\it Ansatz} may reduce the original system of coupled non-linear
partial differential equations to something simpler, and special solutions
may be found by quadratures. So for instance, if the background is axially
symmetric one can look for circular strings, if the background is stationary
one can look for stationary strings etc.

A typical feature of non-linear systems is the presence of regions of unstable
and chaotic motion, i.e. perturbations around certain special solutions develop
imaginary frequencies and grow infinitely. In a recent paper Loust\'{o} and
S\'{a}nchez \cite{san1} considered such string-instabilities in black hole and
de Sitter spacetimes. Their starting point was a method of studying string
solutions in curved spacetimes, originally developed by de Vega and S\'{a}nchez
\cite{veg}. By considering first order string perturbations around the center
of mass motion of the string (which is of course just a
geodesic), Loust\'{o} and
S\'{a}nchez fully analysed the behaviour of the solutions and found the regions
of instability in the 3 cases of Schwarzschild, Reissner-Nordstr\"{o}m and de
Sitter backgrounds.

In this paper we will consider a somewhat similar situation. However, we will
take as the unperturbed string configuration a circular string in the same
3 backgrounds, and we will in fact follow the analysis of Ref.\cite{san1}
very closely.
We will use a method, developed by Frolov and the author \cite{fro} (see also
\cite{guv},\cite{car}),to study covariantly the physical perturbations around
circular strings embedded in the curved spacetimes mentioned above.

This program is mainly motivated by the recent interest in the dynamics of
strings in curved backgrounds and the study of string instabilities
in curved backgrounds: Basu, Guth and Vilenkin
\cite{vil1,vil}
showed that circular cosmic strings may nucleate in the end of the de Sitter
phase of the evolution of the universe (see also \cite{ver}, \cite{axe}) and
thereby avoid to be inflated away from our visible universe. Later Vilenkin
and Garriga \cite{gar1} considered small perturbations around
these nucleated strings both in the end of the de Sitter phase and after
having entered the radiation dominated aera. The evolution of circular cosmic
strings in a radiation dominated universe has also been considered in
\cite{akd,vil2,zha}. Furthermore, circular strings have recently been
discussed in the context of a more systematic investigation of string
dynamics in curved spacetimes \cite{san2,mic,com,ini,all1}, without considering
perturbations around the rings, however. Finally we mention that
superconducting
charge-current carrying circular strings in black hole backgrounds have been
considered in \cite{all2,all3,all4}.

Our work is a natural continuation of the analysis of Loust\'{o} and
S\'{a}nchez \cite{san1} and we hope it will give more insight into the
connection between curved spacetimes and string-instabilities. We will confirm
that in some cases (for instance de Sitter spacetime at
$r\rightarrow\infty$ \cite{ven1,ven2}) the instabilities
are really due to general features of the underlying curved background, while
in other cases (for instance Reissner-Nordstr\"{o}m black hole for
$r\rightarrow 0$) they are just artifacts of the dynamics of the special
unperturbed solution considered.
\vskip 12pt

\hspace*{-6mm}The
paper is organized as follows: In section 2 we will derive the equations
of motion for the unperturbed circular string in the Schwarzschild,
Reissner-Nordstr\"{o}m and de Sitter backgrounds. Then we use the general
formalism of Ref.\cite{fro} to obtain the linearised equations determining
the physical (transverse) perturbations. For simplicity we will take only
4-dimensional spacetimes; it is of course trivial to include more "angular"
space coordinates but concerning the perturbations they will all behave in
the same way in our analysis. For both physical perturbations (one "radial"
and one "angular") we get Schr\"{o}dinger-like equations of the form:
\begin{eqnarray}
&\frac{d^2 f}{d\tau^2}+V(r(\tau))f=0,&\nonumber
\end{eqnarray}
where $f$ is the comoving perturbation, $r(\tau)$ is the radius of
the unperturbed circular string and the string time $\tau$ plays the role
of the spatial coordinate.

In section 3 we analyse these equations, taking $r$ as a parameter. From the
sign of $V(r)$ we find the regions where we expect that the perturbations
develop imaginary frequencies and eventually grow infinitely, in the 3
backgrounds considered. While in section 3 we only get indications of the
emergence of string-instabilities, in section 4 we consider the exact
time-evolution of the perturbations in the regions
$r\rightarrow 0$ (ring-collapse) and $r\rightarrow\infty$. This
provides a connection between the less
strictly obtained results of section 3 and the question of bounded/unbounded
perturbations.
\vskip 12pt

\hspace*{-6mm}The
details of our results are presented in sections 3 and 4, and are
summarized in Fig.2.:

For $r\rightarrow 0$ the perturbations in the direction perpendicular to the
string plane are bounded in all 3 backgrounds, while the perturbations in the
plane of the string grow infinitely.

For $r\rightarrow\infty$ both physical perturbations are boundeed in the case
of Schwarzschild and Reissner-Nordstr\"{o}m black holes (which is not
surprising since these spacetimes are asymptotically flat), but unbounded in
the case of de Sitter spacetime.

\vskip 12pt

\hspace*{-6mm}Throughout
the paper we use sign-conventions of Misner-Thorne-Wheeler \cite{mis}
and units where $G=1$, $c=1$ and the string tension $(2\pi\alpha')^{-1}=1$.
\section{Equations for the string-perturbations}
The classical equations of motion for the bosonic string are in the conformal
gauge given by:
\begin{equation}
\ddot{x}^\mu-x''^\mu+\Gamma^\mu_{\rho\sigma}(\dot{x}^\rho\dot{x}^\sigma-
x'^\rho x'^\sigma)=0,
\end{equation}
where dot and prime denote derivative with respect to the string coordinates
$\tau$ and $\sigma$, respectively. As usual these equations are supplemented
with the 2 gauge constraints:
\begin{equation}
g_{\mu\nu}\dot{x}^\mu x'^\nu=g_{\mu\nu}(\dot{x}^\mu\dot{x}^\nu+x'^\mu x'^\nu)
=0.
\end{equation}
In this paper we will consider perturbations around a circular string
configuration embedded in Schwarzschild, Reissner-Nordstr\"{o}m and de
Sitter spacetimes. In static coordinates these spacetimes are all special
cases of the line element:
\begin{equation}
ds^2=-a(r)dt^2+\frac{dr^2}{a(r)}+r^2d\theta^2+r^2\sin^2\theta d\phi^2,
\end{equation}
so in the first place we will keep $a(r)$ as an arbitrary function. The
components of the Christoffel symbol and of the Riemann tensor (that we
will need later) corresponding to the metric (2.3) are listed in the
appendix.

Let us first consider the unperturbed circular string. It is obtained by
the {\it Ansatz}:
\begin{equation}
t=t(\tau),\;\;r=r(\tau),\;\;\phi=\sigma,\;\;\theta=\pi/2,
\end{equation}
describing a circular string in the equatorial plane with only one physical
mode namely the radial $r(\tau)$. The equations of motion and the constraints
(2.1),(2.2) lead to:
\begin{eqnarray}
&\ddot{t}+\frac{a_{,r}}{a}\dot{t}\dot{r}=0,&\nonumber\\
&\ddot{r}-\frac{a_{,r}}{2a}\dot{r}^2+\frac{1}{2}aa_{,r}\dot{t}^2+
ar=0,&\nonumber\\
&-a\dot{t}^2+\frac{\dot{r}^2}{a}+r^2=0.&
\end{eqnarray}
This system of equations can alternatively be formulated as a Hamiltonian
system:
\begin{equation}
{\cal H}=\frac{1}{2}aP^2_r-\frac{1}{2a}P^2_t+\frac{r^2}{2}\equiv 0.
\end{equation}
We can then eliminate the cyclic coordinate $t$:
\begin{equation}
P_t=-a\dot{t}=const.\equiv -E,\;\;\;\;i.e.:\;\dot{t}=\frac{E}{a},
\end{equation}
where $E$ is the "energy" of the string. The radial coordinate $r$ is then
determined by:
\begin{equation}
\dot{r}^2-E^2+ar^2=0,
\end{equation}
that is solved by:
\begin{equation}
\tau-\tau_o=\pm\int^r_{r_o}\frac{dx}{\sqrt{E^2-x^2a(x)}}.
\end{equation}
By inverting this relation for $r(\tau)$ we then obtain $t(\tau)$ by
integration of equation (2.7).
Note also that the line element (2.3) is now:
\begin{equation}
ds^2=r^2(d\sigma^2-d\tau^2),
\end{equation}
i.e. the invariant string size is given by $r(\tau)$.
\vskip 12pt

\hspace*{-6mm}We will
now consider the perturbations in the 2 physical directions normal
to the string world-sheet. From the 2 tangent vectors:
\begin{equation}
\dot{x}^\mu=(\dot{t},\;\dot{r},\;0,\;0),\;\;x'^\mu=(0,\;0,\;0,\;1),
\end{equation}
we find the 2 normal vectors:
\begin{equation}
n^\mu_\perp=(0,\;0,\;\frac{1}{r},\;0),\;\;n^\mu_\parallel=
(\frac{\dot{r}}{ar},\;\frac{\dot{t}a}{r},\;0,\;0),
\end{equation}
fulfilling the following equations:
\begin{equation}
g_{\mu\nu}n^\mu_R x^\nu_{,A}=0,\;\;g_{\mu\nu}n^\mu_R n^\nu_S=\delta_{RS}.
\end{equation}
Here $(R,S)$ takes the values $"\parallel"$ and $"\perp"$ and
$A=(\tau,\sigma)$. Obviously $n^\mu_\perp$ is perpendicular to
the string plane (that is equal to the equatorial plane) while
$n^\mu_{\parallel}$ is in the string plane. The general physical perturbation
can then be expressed as:
\begin{equation}
\delta x^\mu=n^\mu_{\parallel}\delta x_{\parallel}+n^\mu_{\perp}
\delta x_{\perp},
\end{equation}
where $\delta x_{\parallel}$ and $\delta x_{\perp}$ are the
comoving perturbations, i.e. the perturbations as seen by an observer
travelling with the unperturbed circular string. In the following we will
call these perturbations for the angular perturbation
$(\delta x_{\perp})$ and the radial perturbation
$(\delta x_{\parallel})$, respectively.

According to the general covariant analysis of physical perturbations
propagating along strings in curved spacetimes, carried out by Frolov
and the author \cite{fro},
the perturbations are determined by the following matrix equation:
\begin{eqnarray}
&\Box\delta x_R+2\mu_{RS}\hspace*{.5mm}^A (\delta x^S)_{,A}+(\nabla_A
\mu_{RS}\hspace*{.5mm}^A)\delta x^S
-\mu_{RT}\hspace*{.5mm}^A\mu_S
\hspace*{.2mm}^T\hspace*{.2mm}_A\delta x^S&\nonumber\\
&+\frac{2}{G^C\hspace*{.2mm}_C}\Omega_R\hspace*{.2mm}^{AB}
\Omega_{S,AB}\delta x^S-h^{AB}x^\mu_{,A}x^\nu_{,B}R_{\mu\rho\sigma\nu}
n^\rho_R n^\sigma_S\delta x^S=0.&
\end{eqnarray}
Here $h_{AB}$ and $G_{AB}$ are the intrinsic and induced metric, respectively,
while $\Omega_{R,AB}$ and $\mu_{RS,A}$ are the second fundamental form and
normal fundamental form \cite{eis}, respectively:
\begin{equation}
\Omega_{R,AB}=g_{\mu\nu}n^\mu_R x^\rho_{,A}\nabla_\rho x^\nu_{,B},
\end{equation}
\begin{equation}
\mu_{RS,A}=g_{\mu\nu}n^\mu_R x^\rho_{,A}\nabla_\rho n^\nu_S,
\end{equation}
where $\nabla_\rho$ is the spacetime covariant derivative. $\Box$ and
$\nabla_A$ are the world-sheet d'Alambertian and covariant derivative,
respectively:
\begin{equation}
\Box=\frac{1}{\sqrt{-h}}\partial_A(\sqrt{-h}h^{AB}\partial_B),
\end{equation}
\begin{equation}
\nabla_A\mu_{RS}\hspace*{.5mm}^A=\partial_A\mu_{RS}\hspace*{.5mm}^A+
\Gamma^A_{BA}\mu_{RS}\hspace*{.5mm}^B,\;\;\;\;etc.
\end{equation}
Finally $R_{\mu\rho\sigma\nu}$ represents the spacetime Riemann tensor.

This extremely complicated system of 2 coupled linear second order partial
differential equations fortunately simplifies enormously for the special cases
considered here. One can show that all components of the normal fundamental
form vanish, while the only non-vanishing components of the second
fundamental form are:
\begin{equation}
\Omega_{\parallel\tau\tau}=\Omega_{\parallel\sigma\sigma}=-E.
\end{equation}
The non-vanishing components of the relevant projections of the Riemann tensor
become:
\begin{equation}
h^{AB}x^\mu_{,A}x^\nu_{,B}R_{\mu\rho\sigma\nu}n^\rho_\perp
n^\sigma_\perp=\frac{r}{2}(ra_{,rr}+a_{,r}),
\end{equation}
\begin{equation}
h^{AB}x^\mu_{,A}x^\nu_{,B}R_{\mu\rho\sigma\nu}n^\rho_\parallel
n^\sigma_\parallel=a-1+\frac{r}{2}a_{,r}.
\end{equation}
Finally the d'Alambertian reduces to (conformal gauge) $\Box=\partial^2_\sigma-
\partial^2_\tau$ and $G^A\hspace*{.2mm}_A=h^{AB}G_{AB}=2r^2$. The original
system (2.15) now decouples and leads to the 2 equations:
\begin{eqnarray}
&(\partial^2_\sigma-\partial^2_\tau)\delta x_\perp-\frac{r}{2}
(ra_{,rr}+a_{,r})\delta x_\perp=0,&\nonumber\\
&(\partial^2_\sigma-\partial^2_\tau)\delta x_\parallel-(a-1
+\frac{r}{2}a_{,r}-2\frac{E^2}{r^2})\delta x_\parallel=0.&
\end{eqnarray}
These equations can further be reduced to ordinary differential equations
by Fourier transforming the comoving perturbations:
\begin{eqnarray}
&\delta x_\perp=\tilde{\sum}_{n=-\infty}^{n=+\infty}C_{n\perp}(\tau)
e^{-in\sigma},&\nonumber\\
&\delta x_\parallel=\tilde{\sum}_{n=-\infty}^{n=+\infty}C_{n\parallel}(\tau)
e^{-in\sigma},&
\end{eqnarray}
where $C_{n\perp}=C^*_{-n\perp},\;C_{n\parallel}=C^*_{-n\parallel}$
and the tilde denotes summation for
$\mid n\mid\neq 0,1$ only. The zero modes and the $\mid n\mid=1$ modes are
excluded from the summations since they do not correspond to "real"
perturbations on a circular string \cite{gar1}. They
describe spacetime
translations and rotations that do not change the shape of the string. They
therefore correspond to simply "jumping" from one unperturbed circular string
to another unperturbed circular string. We are then left with the 2
equations $(\mid n\mid\geq 2)$:
\begin{eqnarray}
&\ddot{C}_{n\perp}+(n^2+\frac{r^2}{2}a_{,rr}+
\frac{r}{2}a_{,r})C_{n\perp}=0,&\nonumber\\
&\ddot{C}_{n\parallel}+(n^2+a-1+\frac{r}{2}a_{,r}-
2\frac{E^2}{r^2})C_{n\parallel}=0.&
\end{eqnarray}
Until now we have kept the function $a$ in the line element (2.3) as an
arbitrary function of $r$. As announced in the abstract we will however only
consider the 3 cases of Schwarzschild, Reissner-Nordstr\"{o}m and de Sitter.
These spacetimes are essentially the scalar curvature
flat cases of (2.3) since, from
the appendix, the condition $R=const\equiv K$ is:
\begin{equation}
R=\frac{2}{r^2}(1-a)-4\frac{a_{,r}}{r}-a_{,rr}=K,
\end{equation}
that is integrated to:
\begin{equation}
a(r)=1+\frac{\alpha}{r}+\frac{\beta}{r^2}-\frac{K}{12}r^2,
\end{equation}
where $\alpha$ and $\beta$ are constants. This expression covers the
cosmologically and gravitationally interesting cases of de Sitter
$(\alpha=\beta=0,\;K=12H^2)$, Schwarzschild $(\beta=K=0,\;\alpha=-2M)$
as well as Reissner-Nordstr\"{o}m $(K=0,\;\alpha=-2M,\;\beta=Q^2)$ spacetimes.
In these cases (2.25) leads to:
\begin{eqnarray}
&\ddot{C}_{n\perp}+(n^2+\frac{\alpha}{2r}+2\frac{\beta}{r^2}-
\frac{K}{6}r^2)C_{n\perp}=0,&\nonumber\\
&\ddot{C}_{n\parallel}+(n^2+\frac{\alpha}{2r}-
2\frac{E^2}{r^2}-\frac{K}{6}r^2)C_{n\parallel}=0,&
\end{eqnarray}
and $r(\tau)$ is determined by (2.9):
\begin{equation}
\tau-\tau_o=\pm\int_{r_o}^r\frac{dx}{\sqrt{\frac{K}{12}x^4-x^2-\alpha x
+(E^2-\beta)}}.
\end{equation}
For $K=0$ (the black hole cases) $r(\tau)$ is then a trigonometric function,
while for $K\neq 0$ (de Sitter case) it is generally, but not always, elliptic.
\section{Analysis of string-perturbations}
\setcounter{equation}{0}
In this section we analyse the equations for the perturbations (2.28) taking
$r$ as the parameter. Both equations are Schr\"{o}dinger-like equations
of the form $\ddot{f}+V(r)f=0$, and we will then say that the solutions are
oscillatory in time $\tau$ if $V(r)$ is positive but non-oscillatory in time
$\tau$, developing imaginary frequencies, if $V(r)$ is negative. We are using
the words oscillatory and non-oscillatory in a weak (and sloppy) sense, that
should not be confused with the more strict use of the words in the
mathematical literature, where (say) oscillatory behaviour usually means
that a function has infinitely many zeroes. Note also that there is no
simple one-to-one correspondence between oscillatory (non-oscillatory) and
boundedness (unboundedness) of the solutions. This can for instance be seen
from the simple example:
\begin{eqnarray}
&\ddot{f}+\frac{1}{4\tau^2}f=0.&\nonumber
\end{eqnarray}
In this case we would say that the solutions are oscillatory on the positive
half-axis, but the general solution:
\begin{eqnarray}
&f(\tau)=A\sqrt{\tau}+B\sqrt{\tau}\log{\tau},&\nonumber
\end{eqnarray}
is in fact neither oscillatory in the strict mathematical sense nor is bounded.
In section 4 we will see however, that in most of the cases considered
here non-oscillatory behaviour of the perturbations, developing imaginary
frequencies, will actually lead to unbounded solutions, indicating that the
underlying unperturbed string configuration is unstable, in agreement with
the string instability characterization given by Loust\'{o} and S\'{a}nchez
\cite{san1}.

It should be stressed also that we are still only talking about the
evolution of the comoving string-perturbations. The transformation to the
perturbations as seen by an observer at rest is a highly non-trivial problem,
that to our knowledge has only been done in the case of flat Minkowski space
\cite{gar1}.
In this paper we consider curved spacetimes where everything is of course
somewhat more complicated.

Let us now consider the perturbations in the 3 cases.
\subsection{Schwarzschild black hole}
In this case the unperturbed string solution is (2.29):
\begin{equation}
r(\tau)=M+\sqrt{M^2+E^2}\cos\tau,
\end{equation}
i.e. the string has its maximal radius $r_{max}=M+\sqrt{M^2+E^2}$ at $\tau=0$.
It then contracts through the horizon $r_{hor}=2M$ for $\tau=\arccos
\frac{M}{\sqrt{M^2+E^2}}\in\;]0,\pi/2[\;$, and eventually falls into the
singularity $r=0$ for $\tau=\arccos\frac{-M}
{\sqrt{M^2+E^2}}\in\;]\pi/2,\pi[\;$.
Mathematically speaking it of course continues oscillating but for our
purposes we only consider the proces of collapse from the maximal radius; this
somehow resembles the radial infall of a point particle. The collapse into the
Schwarzschild singularity of course takes infinite coordinate time. This can
be explicitly seen by integrating (2.7) outside the horizon:
\begin{equation}
t=E\tau+2M\log\mid\frac{\tan\frac{\tau}{2}+\delta}{\tan\frac{\tau}{2}-\delta}
\mid,
\end{equation}
where $\delta\equiv (\sqrt{M^2+E^2}-M)/E$.
It follows
that $t(0)=0$ and $t(\arccos\frac{M}{\sqrt{M^2+E^2}})=\infty$.
We note in passing that the expression (3.2) is remarkably simple as compared
to the corresponding relation for the radial infall of a point particle; see
for instance Ref.\cite{mis}. From the
physical point of view we are mostly interested in the circular string and
its perturbations outside the horizon, but considering our whole analysis as
a stability analysis of some special solutions to some non-linear differential
equations, there is no reason not to continue the solutions all the way
through the horizon into the singularity at $r=0$.

The equations determining the modes of $\delta x_\perp$ and $\delta
x_\parallel$, respectively, are:
\begin{eqnarray}
&\ddot{C}_{n\perp}+(n^2-\frac{M}{r})C_{n\perp}=0,&\nonumber\\
&\ddot{C}_{n\parallel}+(n^2-\frac{M}{r}-2\frac{E^2}{r^2})C_{n\parallel}=0.&
\end{eqnarray}
According to the discussion at the beginning of section 3 $\delta x_\perp$ is
oscillatory in time for $r>M/4$, i.e. the first mode ($\mid n\mid=2$) becomes
non-oscillatory at $r=M/4$. The higher modes become non-oscillatory for
smaller and smaller $r$ and for $r=0$ $\delta x_\perp$ is extremely
non-oscillatory (non-oscillatory for all modes).
$\delta x_\parallel$ is oscillatory in time for $r>(M+\sqrt{M^2+32E^2})/8$ and
the picture is then similar to $\delta x_\perp$ for smaller and smaller $r$.
Note that $(M+\sqrt{M^2+32E^2})/8<r_{max}=M+\sqrt{M^2+E^2}$ so that both
$\delta x_\perp$ and $\delta x_\parallel$ are oscillatory when the string is
near its maximal size. $\delta x_\perp$ is also oscillatory at the horizon
$r=2M$ while $\delta x_\parallel$ is oscillatory (non-oscillatory) at the
horizon for $E^2<7M^2$ ($E^2>7M^2$).
\subsection{Reissner-Nordstr\"{o}m black hole}
In this case the unperturbed string solution is (2.29):
\begin{equation}
r(\tau)=M+\sqrt{M^2+E^2-Q^2}\cos\tau;\;\;\;M^2\geq Q^2,
\end{equation}
and we now have to distinguish between 2 different cases namely
$E^2\geq Q^2$ and $E^2<Q^2$.
\vskip 12pt
\hspace*{-6mm}{\bf (i)}, $E^2\geq Q^2$: The
string has its maximal radius $r_{max}=
M+\sqrt{M^2+E^2-Q^2}$ for $\tau=0$. It first contracts through the horizon
$r_+=M+\sqrt{M^2-Q^2}$ for $\tau=\arccos\frac{\sqrt{M^2-Q^2}}
{\sqrt{M^2+E^2-Q^2}}\in\;]0,\pi/2[\;$ and then through the inner horizon
$r_-=M-\sqrt{M^2-Q^2}$ for $\tau=\arccos\frac{-\sqrt{M^2-Q^2}}
{\sqrt{M^2+E^2-Q^2}}\in\;]\pi/2,\pi[\;$. It eventually falls into the
Reissner-Nordstr\"{o}m singularity $r=0$ for $\tau=\arccos\frac{-M}
{\sqrt{M^2+E^2-Q^2}}\in\;]\pi/2,\pi]\;$.
\vskip 12pt
\hspace*{-6mm}{\bf (ii)}, $E^2<Q^2$: The
dynamics of the unperturbed string is similar to
case {\bf(i)} all the way from $r_{max}$ through the 2 horizons, but in this
case the string reaches a {\it minimal} size $r_{min}=M-\sqrt{M^2+E^2-Q^2}>0$
for $\tau=\pi$.
\vskip 12pt
\hspace*{-6mm}The equations determining the modes of $\delta x_\perp$ and
$\delta x_\parallel$, respectively, are:
\begin{eqnarray}
&\ddot{C}_{n\perp}+(n^2-\frac{M}{r}+2\frac{Q^2}{r^2})C_{n\perp}=0,&\nonumber\\
&\ddot{C}_{n\parallel}+(n^2-\frac{M}{r}-2\frac{E^2}{r^2})C_{n\parallel}=0.&
\end{eqnarray}
Formally the $C_{n\parallel}$-equation is identical to
the $C_{n\parallel}$-equation in the
Schwarzschild case, but one should remember that in this case
the charge of course is
present through the expression for $r(\tau)$.

We now find that $\delta x_\perp$ is oscillatory
for all $r$ if $Q^2\geq M^2/32$.
If $Q^2<M^2/32$ the first mode ($\mid n\mid=2$) is non-oscillatory in the
interval $r\in\;](M+\sqrt{M^2-32Q^2})/8,\;(M-\sqrt{M^2-32Q^2})/8[\;$ and
the higher
modes are non-oscillatory in smaller and smaller sub-intervals (for smaller
$Q^2$). These intervals of non-oscillatory behaviour are furthermore located
between the outer and inner horizons, so in both cases {\bf (i)} and {\bf (ii)}
$\delta x_\perp$ is oscillatory at the horizons, and when $r\rightarrow 0$
and $r\rightarrow r_{min}$, respectively.

For the radial perturbations it turns out that the situation is quite
complicated. We find that $\delta x_\parallel$ is oscillatory for
$r>(M+\sqrt{M^2+32E^2})/8\equiv r_{crit}$, i.e. the first mode
$(\mid n\mid=2)$ becomes non-oscillatory for $r=r_{crit}$ and the higher modes
become non-oscillatory for smaller and smaller $r$. Note that
$r_{crit}<r_{max}$ so that $\delta x_\parallel$ is always oscillatory when
the string has its maximal size. The exact location of this critical radius
compared to the 2 horizons and $r_{min}$ (for $E^2<Q^2$) is however very
complicated since we have 2 parameters ($E^2$ and $Q^2$) to play with, so
almost all situations are possible. The result is most easily visualized
by Fig.1. accompanied by the following comments:

In region {\bf (i1)} we have $E^2>Q^2$ and $r_{crit}>r_+$,
so that $\delta x_\parallel$
is non-oscillatory at the (outer) horizon and all the way towards the
singularity $r=0$.

In region {\bf (i2)} we have $E^2>Q^2$ and $r_+>r_{crit}>r_-$, so that
$\delta x_\parallel$ is oscillatory at the horizon but becomes non-oscillatory
before the inner horizon, from which it is non-oscillatory all the way towards
the singularity $r=0$.

In region {\bf (i3)} we have $E^2>Q^2$ and $r_->r_{crit}$,
so that $\delta x_\parallel$
is oscillatory at both horizons but becomes non-oscillatory for
$r\rightarrow 0$.

In {\bf (ii1)} we have $E^2<Q^2$ and $r_+>r_{crit}>r_-$, so that
$\delta x_\parallel$ is oscillatory at the horizon but becomes
non-oscillatory before the inner horizon. It is then non-oscillatory
all the way to $r_{min}$.

In {\bf (ii2)} we have $E^2<Q^2$ and $r_->r_{crit}>r_{min}$, so that
$\delta x_\parallel$ is oscillatory at both horizons but becomes
non-oscillatory for $r\rightarrow r_{min}$.

Finally in {\bf (ii3)} we have $E^2<Q^2$ and $r_{min}>r_{crit}$, so this is a
very interesting
region since $\delta x_\parallel$ is always oscillatory! That possibility
did not exist for $E^2\geq Q^2$.
\subsection{de Sitter spacetime}
In this case $r(\tau)$ is in general given by a Weierstrass elliptic function.
The detailed dynamics of unperturbed circular strings has been discussed
elsewhere \cite{vil1,ver,mic,com,all1} so we
shall not go into it here. We will consider only the
following 3 types of solutions, whose existence is clear from (2.8) when
$a=1-H^2r^2$:
\vskip 12pt
\hspace*{-6mm}{\bf (i)}: For $4H^2E^2\leq 1$ there is a solution
starting with a maximal
radius $r^2_{max}=(1-\sqrt{1-4H^2E^2})/2H^2$ and then collapsing to $r=0$.
It is always inside the horizon $r_{hor}=1/H$.
\vskip 12pt
\hspace*{-6mm}{\bf (ii)}: Still for $4H^2E^2\leq 1$ there is
another solution starting
with a minimal radius $r^2_{min}=(1+\sqrt{1-4H^2E^2})/2H^2$ and then expanding
through the horizon towards infinity.
\vskip 12pt
\hspace*{-6mm}{\bf (iii)}: For $4H^2E^2>1$ there is
a solution starting at $r=0$ and then
expanding through the horizon towards infinity.
\vskip 12pt
\hspace*{-6mm}The equations
for the modes of $\delta x_\perp$ and $\delta x_\parallel$, respectively, are:
\begin{eqnarray}
&\ddot{C}_{n\perp}+(n^2-2H^2r^2)C_{n\perp}=0,&\nonumber\\
&\ddot{C}_{n\parallel}+(n^2-2H^2r^2-2\frac{E^2}{r^2})C_{n\parallel}=0.&
\end{eqnarray}
Let us consider the 3 configurations described above one by one:
\vskip 12pt
\hspace*{-6mm}{\bf (i)}: $\delta x_\perp$
is oscillatory for $r^2<2/H^2$, i.e. it is always
oscillatory since $r^2_{max}<2/H^2$. $\delta x_\parallel$ is oscillatory
in the interval $r^2\in\;](1-\sqrt{1-E^2H^2})/H^2,\;(1+
\sqrt{1-E^2H^2})/H^2[\;$, i.e.
the first mode ($\mid n\mid=2$) becomes non-oscillatory at the boundaries
of this interval. The higher modes are oscillatory in larger and larger
intervals containing the above interval. Note that $(1+\sqrt{1-E^2H^2})/H^2
>r^2_{max}>(1-\sqrt{1-E^2H^2})/H^2$ so that $\delta x_\parallel$ is
always oscillatory when the string has its maximal size. On the other hand
$\delta x_\parallel$ is extremely non-oscillatory for $r\rightarrow 0$.
\vskip 12pt
\hspace*{-6mm}{\bf (ii)}: The critical string radii
are the same as in case {\bf (i)}, but
the dynamics of the unperturbed string is completely different. We have
that $r^2_{min}<2/H^2$ so that $\delta x_\perp$ is oscillatory at $r_{min}$
and at the horizon, but it is non-oscillatory from $r^2=2/H^2$ towards
infinity. For $\delta x_\parallel$ the picture is more or less the same. We
find that $(1+\sqrt{1-E^2H^2})/H^2>r^2_{hor}>r^2_{min}
>(1-\sqrt{1-E^2H^2})/H^2$ so that $\delta x_\parallel$ is oscillatory at
$r_{min}$ and at the horizon, but it is non-oscillatory from
$r^2=(1+\sqrt{1-E^2H^2})/H^2$ towards infinity. Note that the non-oscillatory
behaviour of $\delta x_\parallel$ sets in a little before $\delta x_\perp$,
but in both cases it is outside the horizon, and of course for higher and
higher modes the non-oscillatory behaviour sets in for larger and larger $r$.
\vskip 12pt
\hspace*{-6mm}{\bf (iii)}: Again the
critical radii are the same as in the other 2 cases.
$\delta x_\perp$ is oscillatory from $r=0$ through the horizon to $r^2=2/H^2$
where the first ($\mid n\mid=2$) non-oscillatory behaviour sets in. It is
then non-oscillatory all the way towards infinity. The higher modes become
non-oscillatory for larger $r$. Finally $\delta x_\parallel$ is oscillatory
in the interval $r^2\in\;](1-\sqrt{1-E^2H^2})/H^2,\;(1+\sqrt{1-E^2H^2})/H^2[\;$
surrounding the horizon, but it is extremely non-oscillatory for both
$r\rightarrow 0$ and $r\rightarrow\infty$.
\vskip 12pt
\hspace*{-6mm}In de Sitter space there is also a stationary circular string
solution \cite{fro,mic,com}. From (2.8) follows that
it is described by $4E^2H^2=1,\;2H^2r^2=1$.
The stability of this solution was already considered in \cite{fro} in a
different gauge, so let us restate the result here. The equations for the
modes of $\delta x_\perp$ and $\delta x_\parallel$, respectively, are (3.6):
\begin{eqnarray}
&\ddot{C}_{n\perp}+(n^2-1)C_{n\perp}=0,&\nonumber\\
&\ddot{C}_{n\parallel}+(n^2-2)C_{n\parallel}=0.&
\end{eqnarray}
So, as already stated in \cite{fro}, we get instabilities for the
$\mid n\mid=1$ modes (and for the zero modes). However, as explained after
equation (2.24), we do not consider these modes as "real" perturbations
\cite{gar1} since they
just correspond to "jumping" from one unperturbed
string to another one, without changing shape. It may be a matter of taste
whether or not to include the $\mid n\mid=1$ modes, but in any case we get
the somewhat surprising result that the stationary circular string solution
in de Sitter spacetime is actually stable against "real" perturbations where
$\mid n\mid\geq 2$.
\vskip 12pt
\hspace*{-6mm}This concludes
our investigations of oscillatory and non-oscillatory (in the
weak sense considered here) behaviour of the comoving perturbations around
a circular string in the 3 backgrounds of Schwarzschild,
Reissner-Nordstr\"{o}m and de Sitter spacetimes. In the next section we
will relate some of these results to the question of bounded or unbounded
comoving perturbations.
\section{Time-evolution and asymptotic behaviour}
\setcounter{equation}{0}
In this section we address the question of bounded or unbounded comoving
perturbations by considering the time evolution of some of the solutions
found in section 3, i.e. we now take $\tau$ as the parameter. It is clear from
the general equations determining the perturbations (2.28) that we can only
expect unbounded behaviour of the solutions in the 2 regions $r\rightarrow 0$
and $r\rightarrow\infty$, so we will restrict ourselves by considering the
solutions in these regions only. It is also clear that in the 2 cases of black
holes we need only consider $r\rightarrow 0$ since for $r\rightarrow\infty$
(that can of course only be obtained for $E^2\rightarrow\infty$, i.e.
infinite string energy) we have Minkowski space, where the perturbations are
obviously bounded; they are just ordinary plane waves. For de Sitter
spacetime, on the other hand, we can get unbounded perturbations for both
$r\rightarrow 0$ and $r\rightarrow\infty$.

In all cases it will turn out that the behaviour of the
perturbations in the asymptotic regions corresponds to different cases
of the motion of a particle in the potential $\alpha(\tau-\tau_o)^{-\beta}$
\cite{san1,san3,med}, in the sense that they are described by a
stationary Schr\"{o}dinger
equation with $\tau$ playing the role of the spatial
parameter. It is an elementary observation that if $\alpha<0$ and
$\beta\geq 2$ there are singular solutions for $\tau\rightarrow\tau_o$.
Therefore, as soon as we have obtained the potential with the 2 parameters
$\alpha$ and $\beta$ we can conclude whether the perturbations
blow up, indicating that the underlying circular string is unstable.
For completeness we will however give the full solutions in the
asymptotic regions, demonstrating explicitly if and how the perturbations
blow up.
\subsection{Schwarzschild black hole}
In this case we have $r(\tau)=M+\sqrt{M^2+E^2}\cos\tau$ with $r\rightarrow 0$
corresponding to $\tau\rightarrow\tau_{o}\equiv\arccos\frac{-M}{\sqrt
{M^2+E^2}}$ from below (cf. subsection 3.1). For $r\rightarrow 0$ we then have
approximately:
\begin{equation}
r(\tau)\approx\mid E\mid(\tau-\tau_o),
\end{equation}
and the 2 equations determining the perturbations (3.3) become approximately:
\begin{eqnarray}
&\ddot{C}_{n\perp}-\frac{M}{\mid E\mid (\tau_o-\tau)}C_{n\perp}=0,&\nonumber\\
&\ddot{C}_{n\parallel}-\frac{2}{(\tau_o-\tau)^2}C_{n\parallel}=0.&
\end{eqnarray}
Let us first consider the perturbations in the angular direction
(the $C_{n\perp}$'s).
Keeping in mind that $\tau_o-\tau$ is positive in the relevant range of $\tau$
we find the 2 real independent solutions in terms of Bessel functions
\cite{abr}:
\begin{equation}
f=\sqrt{\tau_o-\tau}J_1(2\sqrt{M/\mid E\mid}\sqrt{\tau_o-\tau}),\;\;\;
g=\sqrt{\tau_o-\tau}N_1(2\sqrt{M/\mid E\mid}\sqrt{\tau_o-\tau}).
\end{equation}
The most interesting feature of these solutions is that they are actually
bounded \cite{abr}:
\begin{equation}
f\rightarrow\sqrt{M/\mid E\mid}(\tau_o-\tau),\;\;\;g\rightarrow-\frac{1}{\pi}
\sqrt{\mid E\mid/ M};\;\;\;\;\tau\rightarrow\tau_o.
\end{equation}
This therefore provides an example where the solutions were classified as
non-oscillatory (according to section 3.1), but where the actual time evolution
demonstrates that the solutions are bounded, and they are in fact oscillatory
in the strict mathematical sense of having infinitely many zeroes. This is
however an exceptional case; in the other cases under consideration here we
will find that non-oscillatory behaviour at $r\rightarrow 0$ or
$r\rightarrow\infty$ leads to unbounded solutions.

For the perturbations in the radial direction (the $C_{n\parallel}$'s) we find
the
complete solution:
\begin{equation}
C_{n\parallel}(\tau)=\alpha_n(\tau_o-\tau)^2+\frac{\beta_n}{\tau_o-\tau},
\end{equation}
where $\alpha_n$ and $\beta_n$ are arbitrary constants. This solution
is indeed unbounded for $\tau\rightarrow\tau_o$.
\subsection{Reissner-Nordstr\"{o}m black hole}
Here we only consider the case where $E^2\geq Q^2$ to ensure that we have
solutions collapsing to $r=0$. Then the unperturbed string is determined by
$r(\tau)=M+\sqrt{M^2+E^2-Q^2}\cos\tau$ and $r\rightarrow 0$ corresponds to
$\tau\rightarrow\tau_o\equiv\arccos\frac{-M}{\sqrt{M^2+E^2-Q^2}}$ from
below (cf. subsection 3.2). The approximate solution for $r\rightarrow 0$ is
then:
\begin{equation}
r(\tau)\approx\sqrt{E^2-Q^2}(\tau_o-\tau);\;\;\;\;E^2>Q^2
\end{equation}
and:
\begin{equation}
r(\tau)\approx\frac{M}{2}(\tau_o-\tau)^2;\;\;\;\;E^2=Q^2
\end{equation}
The 2 equations determining the perturbations (3.5) become:
\begin{eqnarray}
&\ddot{C}_{n\perp}+\frac{2Q^2}{(E^2-Q^2)(\tau_o-\tau)^2}
C_{n\perp}=0;&\;\;\;\;E^2>Q^2\nonumber\\
&\ddot{C}_{n\parallel}-\frac{2E^2}{(E^2-Q^2)(\tau_o-\tau)^2}
C_{n\parallel}=0;&\;\;\;\;E^2>Q^2
\end{eqnarray}
and:
\begin{eqnarray}
&\ddot{C}_{n\perp}+\frac{8Q^2}{M^2(\tau_0-\tau)^4}
C_{n\perp}=0;&\;\;\;\;E^2=Q^2\nonumber\\
&\ddot{C}_{n\parallel}-\frac{8Q^2}{M^2(\tau_o-\tau)^4}
C_{n\parallel}=0;&\;\;\;\;E^2=Q^2
\end{eqnarray}

The solution of (4.8) is:
\begin{eqnarray}
&C_{n\perp}(\tau)=\alpha_n(\tau_o-\tau)^{(1+\rho)/2}+\beta_n(\tau_o-\tau)^
{(1-\rho)/2},&\nonumber\\
&C_{n\parallel}(\tau)=
\gamma_n(\tau_o-\tau)^{(1+\eta)/2}+\delta_n(\tau_o-\tau)^
{(1-\eta)/2},&
\end{eqnarray}
where $\rho=\sqrt{1-\frac{8Q^2}{E^2-Q^2}},\;\eta=\sqrt{1+\frac{8E^2}{E^2-Q^2}}$
and $(\alpha_n, \beta_n, \gamma_n, \delta_n)$ are arbitrary constants. These
constants should be chosen such that $C_{n\perp}(\tau)$
and $C_{n\parallel}(\tau)$ are real
functions; note that $\rho$ is not necessarily real. In any case
we find that $C_{n\perp}(\tau)$ is finite for $\tau\rightarrow\tau_o$
while $C_{n\parallel}(\tau)$
blows up because of the term multiplied by $\delta_n$. The expression for
$C_{n\perp}(\tau)$ in (4.10) is actually only the general solution provided
$E^2\neq 9Q^2$. For $E^2=9Q^2$ we find instead:
\begin{equation}
C_{n\perp}(\tau)=\alpha_n\sqrt{\tau_o-\tau}+\beta_n\sqrt{\tau_o-\tau}
\log{(\tau_o-\tau)},
\end{equation}
which is however also bounded for $\tau\rightarrow\tau_o$.

For $E^2=Q^2$ (4.9) the general solution is:
\begin{eqnarray}
&C_{n\perp}(\tau)=
\alpha_n(\tau_o-\tau)\sin\frac{\sqrt{8}\mid Q/M\mid}{\tau_o-\tau}+
\beta_n(\tau_o-\tau)\cos\frac{\sqrt{8}\mid Q/M\mid}{\tau_o-\tau},&\nonumber\\
&C_{n\parallel}(\tau)=
\gamma_n(\tau_o-\tau)\sinh\frac{\sqrt{8}\mid Q/M\mid}{\tau_o-\tau}+
\delta_n(\tau_o-\tau)\cosh\frac{\sqrt{8}\mid Q/M\mid}{\tau_o-\tau},&
\end{eqnarray}
where $(\alpha_n, \beta_n, \gamma_n, \delta_n)$ are arbitrary real constants.
So again we find that $C_{n\perp}(\tau)$ is
finite for $\tau\rightarrow\tau_o$ while
$C_{n\parallel}(\tau)$ blows up.

The results of the analysis of this subsection confirm what we found in
subsection 3.2: Non-oscillatory behaviour in the region $r\rightarrow 0$ leads
to unbounded solutions, while oscillatory behaviour in the regions
$r\rightarrow 0$ and $r\rightarrow\infty$ leads to bounded (finite)
solutions.
\subsection{de Sitter spacetime}
Finally, we come to the de Sitter case, and we first consider the asymptotic
region $r\rightarrow\infty$, i.e. we consider the strings {\bf (ii)}
and {\bf (iii)}
of subsection 3.3. The asymptotic behaviour of the unperturbed string is
most easily found directly from the equation of motion (2.8). For
$a=1-H^2r^2$ and $r\rightarrow\infty$ we find:
\begin{equation}
\dot{r}^2-H^2r^4\approx 0,
\end{equation}
so that for an expanding solution $(\dot{r}\geq 0)$:
\begin{equation}
r(\tau)\approx\frac{1}{H(\tau_o-\tau)},
\end{equation}
for some constant $\tau_o$, and $r\rightarrow\infty$ corresponds to
$\tau\rightarrow\tau_o$ from below. The 2 equations determining the
perturbations (3.6) become:
\begin{eqnarray}
&\ddot{C}_{n\perp}-\frac{2}{(\tau_o-\tau)^2}C_{n\perp}=0,&\nonumber\\
&\ddot{C}_{n\parallel}-\frac{2}{(\tau_o-\tau)^2}C_{n\parallel}=0.&
\end{eqnarray}
It follows that they both blow up in exactly
the same way as $C_{n\parallel}(\tau)$ blows up
at $r\rightarrow 0$ in the Schwarzschild case; compare with (4.2),(4.5).

In the region $r\rightarrow 0$
(considering now the unperturbed strings {\bf (i)}
and {\bf (ii)} of subsection 3.3) we find from (2.8) the asymptotic
behaviour $\dot{r}^2\approx E^2$. For a collapsing string $(\dot{r}\leq 0)$
this leads to:
\begin{equation}
r(\tau)\approx\mid E\mid(\tau_o-\tau),
\end{equation}
for some constant $\tau_o$, and $r\rightarrow 0$ corresponds to
$\tau\rightarrow\tau_o$ from below. The 2 equations determining the
perturbations (3.6) in this limit become:
\begin{eqnarray}
&\ddot{C}_{n\perp}+n^2C_{n\perp}=0,&\nonumber\\
&\ddot{C}_{n\parallel}-\frac{2}{(\tau_o-\tau)^2}C_{n\parallel}=0.&
\end{eqnarray}
Obviously $C_{n\perp}(\tau)$ is finite while $C_{n\parallel}(\tau)$ blows
up in the same way as for
$r\rightarrow\infty$.

As in the Reissner-Nordstr\"{o}m case the results obtained in this subsection
confirm what we found in section 3.
\vskip 12pt
\section{Conclusion}
In conclusion we have studied the comoving perturbations around circular
strings embedded in the curved spacetimes of Schwarzschild,
Reissner-Nordstr\"{o}m and de Sitter.
The results of our analysis are summarized in Fig.2., the details are
presented in sections 3 and 4. The main conclusions were already drawn
in the introduction but let us say a few more words here. Our results for
the perturbations on the circular strings in de Sitter spacetime in the
asymptotic region $r\rightarrow\infty$ confirm the results of Loust\'{o}
and S\'{a}nchez \cite{san1} and also
the results of Gasperini, S\'{a}nchez and Veneziano
\cite{ven1,ven2} for highly unstable strings.  For the
black holes we get the same results as Loust\'{o}
and S\'{a}nchez in the Schwarzschild case, while in the Reissner-Nordstr\"{o}m
case the results are different for the angular perturbations. This is however
a "normal" situation; some special solutions of a non-linear differential
equation are stable and some are unstable. In this sense the de Sitter
spacetime at $r\rightarrow\infty$ provides an exceptional case.
\vskip 24pt
\hspace*{-6mm}{\bf Acknowledgements:} I would like to thank the Observatoire
de Paris, Section de Meudon, Demirm, where this work was completed,
for kind hospitality. I especially thank N. S\'{a}nchez for discussions and
suggestions to improve the original manuscript.
\newpage
\section{Appendix}
\setcounter{equation}{0}
In this appendix we give the explicit expressions for the non-vanishing
components of the Christoffel symbol and Riemann tensor corresponding to
the line element (2.3).
\vskip 12pt
\hspace*{-6mm}{\bf The metric:}
\begin{equation}
g_{tt}=-a,\;\;g_{rr}=\frac{1}{a},\;\;g_{\theta\theta}=r^2,\;\;g_{\phi\phi}=
r^2\sin^2\theta;\;\;\;\;a=a(r).
\end{equation}
\vskip 12pt
\hspace*{-6mm}{\bf The Christoffel symbol:}
\begin{eqnarray}
&\Gamma^t_{tr}=\frac{a_{,r}}{2a},\;\;\Gamma^r_{rr}=\frac{-a_{,r}}{2a},\;\;
\Gamma^r_{tt}=\frac{1}{2}aa_{,r},&\nonumber\\
&\Gamma^r_{\theta\theta}=-ar,\;\;\Gamma^r_{\phi\phi}=-ar\sin^2\theta,\;\;
\Gamma^\phi_{\phi r}=\frac{1}{r},&\nonumber\\
&\Gamma^\phi_{\phi\theta}=\cot\theta,\;\;\Gamma^\theta_{\theta r}=\frac{1}
{r},\;\;\Gamma^\theta_{\phi\phi}=-\sin\theta\cos\theta.&
\end{eqnarray}
\vskip 12pt
\hspace*{-6mm}{\bf The Riemann tensor:}
\begin{eqnarray}
&R_{rtrt}=\frac{1}{2}a_{,rr},\;\;R_{r\theta r\theta}=\frac{-r}{2a}a_{,r},\;\;
R_{r\phi r\phi}=\frac{-r}{2a}a_{,r}\sin^2\theta,&\nonumber\\
&R_{t\theta t\theta}=\frac{r}{2}aa_{,r},\;\;R_{t\phi t\phi}=\frac{r}{2}a
a_{,r}\sin^2\theta,\;\;R_{\theta\phi\theta\phi}=r^2(1-a)\sin^2\theta.&
\end{eqnarray}
\vskip 12pt
\hspace*{-6mm}{\bf The Ricci tensor:}
\begin{equation}
R_{tt}=-a^2R_{rr}=a(\frac{a_{,rr}}{2}+\frac{a_{,r}}{r}),\;\;
R_{\phi\phi}=\sin^2\theta R_{\theta\theta}=(1-a-ra_{,r})\sin^2\theta.
\end{equation}
\vskip 12pt
\hspace*{-6mm}Finally the {\bf scalar curvature} is:
\begin{equation}
R=-a_{,rr}+\frac{2}{r^2}(1-a)-\frac{4}{r}a_{,r}.
\end{equation}

\newpage
\begin{center}
\bf{Figure Captions}
\end{center}
\vskip 12pt
Fig.1. The location of the critical radius where the
non-oscillatory behaviour sets in for the $\mid n\mid=2$ mode in the case
of Reissner-Nordstr\"{o}m black hole. The details
of this figure are explained at the end of subsection 3.2. Note that we only
consider $Q^2\leq M^2$.
\vskip 36pt
\hspace*{-6mm}Fig.2. This diagram summarizes the results obtained in this
paper. $\delta x_\perp$ and $\delta x_\parallel$ corresponds to the
comoving angular and radial perturbations, respectively.
\end{document}